\documentclass{article}
\usepackage{spconf}
\usepackage{amsmath,graphicx}
\usepackage{subcaption}
\usepackage[hidelinks]{hyperref}
\usepackage{xcolor}
\usepackage{easyReview}
\usepackage[super]{nth}
\usepackage[detect-weight,retain-explicit-plus]{siunitx}
\usepackage{newtxmath}
\usepackage[scr=rsfso]{mathalfa}
\usepackage{multirow}
\usepackage[subtle]{savetrees}
\usepackage{balance}
\usepackage{hhline}
\usepackage{makecell}
\usepackage[shortlabels]{enumitem}
\usepackage{tabularx}

\usepackage{booktabs}
\usepackage{siunitx}
\usepackage{tikz}
\usetikzlibrary{arrows.meta,positioning,calc}
\usepackage{pgf}

\title{Bin2Ambi: Learning Ambisonic Soundfield Reconstruction from Head-Tracked Binaural Audio}
\name{Gavin Milner, Nils Peters\thanks{Corresponding author: nils.peters@tcd.ie}}
\address{Trinity College Dublin, Dept. of Electronic \& Electrical Engineering, Dublin, Ireland}

\begin{document}
\ninept
\maketitle
\begin{abstract}

User-generated content has become one of the most-consumed content types. However, capturing spatial audio with consumer hardware is still challenging. 
Given the widespread success of smart earbuds,  binaural audio could be a promising option to capture spatial audio on consumer devices, but its inherent signal characteristic limits its usability as a recording format.  
In this paper, we propose and define a new task, Binaural to Ambisonics conversion (Bin2Ambi). In our proposed system, we exploit simultaneously captured head-tracking data provided from the motion sensors in smart earbuds. We show that this motion data help resolve the inherent directional uncertainty of two-channel binaural audio due to front-back localization ambiguities and lateral errors in the cone of confusion.
Our results show that our system learns directional and diffuse-field information and that head-tracking especially reduces extreme localization errors.
Objective metrics and a subjective listening test suggest that the converted Ambisonics soundfield achieves an average directional error of up to $11.8^\circ$ and a perceived spatial quality similar to a DirAC ground-truth model. The proposed algorithm can serve as a baseline for future improvements to this novel Bin2Ambi task.

\end{abstract}

\section{Introduction} \label{sec:intro}
Spatial audio is a growing sound technology creating immersive, realistic auditory experiences by reproducing sound in all directions. It is increasingly used in music, film, gaming, and AR/VR, with further applications in hearing aids, teleconferencing, and the automotive industry. Binaural audio recordings mimic human hearing by using two in-ear microphones, accounting for how the pinna, head, and torso filter sound. 
Binaural audio is intended for headphone playback, limiting its direct application in next-generation audio systems that support object-based and scene-based audio formats (e.g., \cite{MPEG-H_2017,multrus2024immersive}). Ambisonics is a playback-agnostic scene-based audio representation that uses spherical harmonics (SH) to describe a 3D soundfield \cite{olivieri2019scene} and can be rendered to headphones and arbitrary loudspeaker configurations.
To our knowledge, no processing method exists to algorithmically transform a two-channel binaural recording to a four-channel First-Order Ambisonics representation - an inherently underdetermined problem.

In this paper, we introduce the Binaural to Ambisonics conversion task (Bin2Ambi) in which a two-channel binaural audio signal is transformed into a periphonic Ambisonics soundfield representation. 
This task differs from traditional channel-based upmixing (e.g., stereo-to-7.1.4 as in \cite{liang2026immersiveflow}): First, in contrast to stereo, the binaural audio signal is a spatial format, defined by its characteristic interchannel level and temporal cues. Second, the target signal is not a fixed channel-based format, but scene-based Ambisonics. Finally, dynamic head motion inherently affects the binaural signal, which requires motion compensation to generate a stable target signal.

Enabling Bin2Ambi conversion would increase binaural audio's versatility, enabling modern earphones to become low-barrier spatial recording devices for user-generated content.

\subsection{State of the Art}

Binaural localization primarily relies on interaural time, level and phase differences between the two ears. However, these cues can become ambiguous for sources located within the cone of confusion (CoC), particularly for front-back localization. Head movement provides additional dynamic binaural cues which can help resolve these ambiguities. Fleischhauer and Jax \cite{fleischhauerFullSphereBinauralDirectionofArrival2025} demonstrated improved DoA estimation using head movement, with yaw particularly effective at reducing front--back confusion. More recently, learning-based approaches have been used to model these relationships directly from binaural features. García-Barrios et al. \cite{garcia2022binaural} used a CRNN to combine spectral binaural features with quaternion head rotation information for DoA estimation, finding that processing the rotation features on a separate convolutional branch before concatenation gave the best localization performance.

CRNN architectures have been applied to sound source localization by combining convolutional spectral feature extraction with recurrent modelling of temporal information \cite{adavanneSoundEventLocalization2019}, while U-Net architectures have been used to estimate time-frequency spatial parameters for stereo upmixing \cite{Turner2024TFSpatialUpmix}. Existing neural spatial audio conversion methods have primarily considered Ambisonics-to-binaural rendering \cite{yinEndtoEndPairedAmbisonicBinaural2024, gebruA2BNeuralRendering2025}, rather than the inverse problem considered here. 

In this contribution, we propose to estimate the spatial parameters of the DirAC model \cite{pulkkiFirstOrderDirAC2017} from binaural audio features and head-tracking data. We evaluate the proposed neural model across increasing scene complexity and background noise conditions, and investigate the contribution of head rotation to full-spherical localization. Our evaluations comprise objective metrics and a subjective listening test.  

\section{Proposed Architecture} \label{sec:methodology}
\begin{figure}[htbp]
    \centering
    \resizebox{1\columnwidth}{!}
    {\definecolor{blueColour}{HTML}{B0E3E6}
\definecolor{rotColor}{HTML}{E1D5E7}
\definecolor{fuseColor}{HTML}{FFF2CC}
\definecolor{concatColor}{HTML}{999999}
\definecolor{greenColour}{HTML}{D5E8D4}

\begin{tikzpicture}[
    >={Stealth[scale=1.2]},
    thick,
    font=\large,
    textnode/.style={align=center, font=\sffamily\large},
    bluebox/.style={rounded corners=3pt,draw=green!60!black, fill=greenColour!80, minimum width=2.5cm, minimum height=2.4cm, align=center, font=\sffamily\large},
    greenbox/.style={rounded corners=3pt,draw=blueColour!50!black, fill=blueColour!80, minimum width=2.5cm, minimum height=2.4cm, align=center, font=\sffamily\large},
    orangebox/.style={rounded corners=3pt,draw=orange!70!black, fill=orange!20, minimum width=2.5cm, minimum height=2.4cm, align=center, font=\sffamily\large}
]

\node[textnode] (input) at (0, 0) {Binaural Audio (2 ch.)\\+\\Head-Tracking Data};
\node[bluebox, right=0.8cm of input] (binfeat) {Binaural\\Feature\\Extraction};
\node[greenbox, right=3.7cm of binfeat] (dnn) {Neural\\Net};
\node[orangebox, right=3.0cm of dnn] (dirac) {Ambisonics\\Synthesis};
\node[textnode, right=0.8cm of dirac] (output) {First-Order\\Ambisonics\\(4 ch.)};

\draw[->, line width=0.5mm] (input.east) -- (binfeat.west);
\draw[->, line width=0.5mm] (binfeat.east) -- node[text centered, text width=3.5cm,midway,above] {Binaural + Rotation\\Features} (dnn.west);
\draw[->, line width=0.5mm] (dnn.east) -- node[text centered, text  width=2.0cm,midway,above] {DirAC\\Parameter}  (dirac.west);
\draw[->, line width=0.5mm] (dirac.east) -- (output.west);

\draw[->, line width=0.5mm] (binfeat.south) -- ++(0,-0.6) -| node[pos=0.25, below] {Sound Pressure Estimate} (dirac.south);

\end{tikzpicture}}    
    \caption{Proposed architecture for the Bin2Ambi task.}
    \label{fig:inference}
\end{figure}
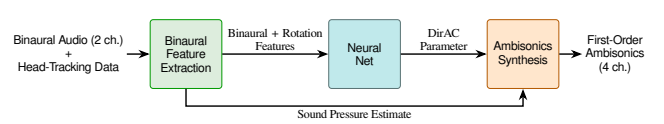

We propose a resource-efficient hybrid DSP/Deep Learning approach as depicted in Fig.~\ref{fig:inference}. The system takes binaural audio signals and corresponding head movement data from a head tracker as input, from which a set of binaural features is extracted. 
These features are fed to a neural network architecture  which estimates for each time-frequency (TF) bin the corresponding directional and diffuseness information of the soundfield as defined by the parametric DirAC model \cite{pulkkiFirstOrderDirAC2017}. Finally, a DirAC decoder synthesizes Ambisonics signals. In this study, we synthesize First-Order Ambisonics (FOA).
Head movement information is useful for two reasons: First, it should help resolve directional ambiguities where similar interaural cues can correspond to multiple source directions, i.e., front/back localization errors and the cone of confusion (CoC) \cite{blauert1996spatial}. Second, head tracker data are needed to remove undesired sound field rotations caused by head movements from the converted Ambisonics signals.

\subsection{Binaural Feature Extraction}

Five frequency-dependent binaural features are extracted from the STFT of the left and right ear signals, $L(f,t), R(f,t) \in \mathbb{C}$, computed at time frame $t$ and frequency bin $f$ using a 1024-sample Hann window with 50\% overlap at \SI{48}{kHz}.
The feature set consists of the Mean Magnitude Spectrogram, Interaural Level Difference (ILD, Eq.~\ref{eq:feat_ild}), Sine and Cosine of Interaural Phase Difference (IPD) to avoid discontinuities caused by phase wrapping (Eqs.~\ref{eq:feat_sin_ipd}-\ref{eq:feat_cos_ipd}), and Interaural Coherence (IC, Eq.~\ref{eq:feat_ic}) where $\epsilon$ is a small constant for robustness and $\Phi_{LR}(f,t) = L(f,t) R^*(f,t)$ is the complex cross-spectrum between the binaural channels, with $(\cdot)^*$ denoting complex conjugation. The symbol $\bar{\Phi}$ denotes recursive temporal smoothing. 

    \begin{equation}
        \mathrm{ILD}(f,t) = 20 \log_{10} \left( \frac{|L(f,t)| + \epsilon}{|R(f,t)| + \epsilon} \right),
        \label{eq:feat_ild}
    \end{equation}

\begin{align}    
        \mathrm{IPD}_{\sin}(f,t) &= \sin\left(\angle \Phi_{LR}(f,t)\right) = \frac{\operatorname{Im}\{\Phi_{LR}(f,t)\}}{|\Phi_{LR}(f,t)| + \epsilon}, \label{eq:feat_sin_ipd} \\
        \mathrm{IPD}_{\cos}(f,t) &= \cos\left(\angle \Phi_{LR}(f,t)\right) = \frac{\operatorname{Re}\{\Phi_{LR}(f,t)\}}{|\Phi_{LR}(f,t)| + \epsilon}, \label{eq:feat_cos_ipd}
\end{align}

    \begin{equation}
    \mathrm{IC}(f,t) =
    \frac{|\bar{\Phi}_{LR}(f,t)|}
    {\sqrt{\bar{\Phi}_{LL}(f,t)\bar{\Phi}_{RR}(f,t)}+\epsilon}
    \label{eq:feat_ic}
\end{equation}

All five features are compressed along the frequency axis using a 64-band Mel-scale triangular filterbank to reduce dimensionality while preserving perceptual frequency resolution, yielding the spectral feature tensor $\mathbf{X} \in \mathbb{R}^{5 \times 64 \times T}$. Simultaneously, head orientation data sampled at \SI{100}{Hz} is synchronized to the STFT frame rate and represented as normalized unit quaternions $\mathbf{Q} \in \mathbb{R}^{4 \times T}$, with $\mathbf{q}(t) = [q_w(t), q_x(t), q_y(t), q_z(t)]^T$ following \cite{garcia2022binaural}.

\subsection{Neural Network Architecture}\label{sec:model}

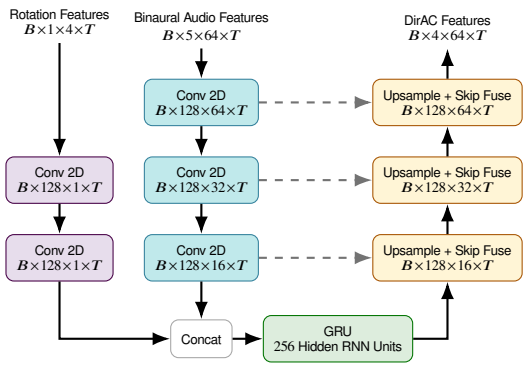
\begin{figure}[htbp]
    \centering
    \definecolor{convColor}{HTML}{B0E3E6}
\definecolor{rotColor}{HTML}{E1D5E7}
\definecolor{fuseColor}{HTML}{FFF2CC}
\definecolor{concatColor}{HTML}{999999}
\definecolor{gruColor}{HTML}{D5E8D4}

\begin{tikzpicture}[
  font=\sffamily\tiny,
  >=Latex,
  node distance=4mm and 4mm,
  block/.style={draw, rounded corners=3pt, align=center,
    minimum width=5mm, minimum height=5mm, inner sep=4pt},
  conv/.style={block, fill=convColor!80, draw=cyan!55!black},
  rot/.style={block, fill=rotColor!80, draw=violet!55!black,
    minimum width=5mm},
  fuse/.style={block, fill=fuseColor!80, draw=orange!65!black},
  concat/.style={block, fill=white!25, draw=white!65!black},
  gru/.style={block, fill=gruColor!80, draw=green!45!black,
    minimum width=15mm},
  flow/.style={->, thick},
  skip/.style={->, dashed, thick, black!55}
]

\node (spec) {\shortstack{Binaural Audio Features \\$B\times5\times64\times T$}};
\node[conv, below=of spec] (enc1) {Conv 2D\\$B\times128\times64\times T$};
\node[conv, below=of enc1] (enc2) {Conv 2D\\$B\times128\times32\times T$};
\node[conv, below=of enc2] (enc3) {Conv 2D\\$B\times128\times16\times T$};


\node[rot, left=of enc2] (rot1) {Conv 2D\\$B\times128\times1\times T$};
\node[rot, below=of rot1] (rot2) {Conv 2D\\$B\times128\times1\times T$};
\node[above=1.5cm of rot1, align=center] (rotin) {Rotation Features\\$B\times1\times4\times T$};
\node[concat, below=0.5cm of enc3] (concat) {Concat};

\node[gru, right=of concat] (gru) {GRU\\ $256$ Hidden RNN Units};

\node[fuse, right=15mm of enc3] (dec3) {Upsample + Skip Fuse\\$B\times128\times16\times T$};
\node[fuse, above=of dec3] (dec2) {Upsample + Skip Fuse\\$B\times128\times32\times T$};
\node[fuse, above=of dec2] (dec1) {Upsample + Skip Fuse\\$B\times128\times64\times T$};
\node[above=of dec1, align=center] (out) {DirAC Features\\$B\times4\times64\times T$};

\draw[flow] (spec) -- (enc1);
\draw[flow] (enc1) -- (enc2);
\draw[flow] (enc2) -- (enc3);
\draw[flow] (enc3) -- (concat);
\draw[flow] (concat) -- (gru);
\draw[flow] (gru.east) -| (dec3);
\draw[flow] (dec3) -- (dec2);
\draw[flow] (dec2) -- (dec1);
\draw[flow] (dec1) -- (out);

\draw[flow] (rotin) -- (rot1);
\draw[flow] (rot1) -- (rot2);
\draw[flow] (rot2) |- (concat);

\draw[skip] (enc1.east) -- (dec1.west);
\draw[skip] (enc2.east) -- (dec2.west);
\draw[skip] (enc3.east) -- (dec3.west);

\end{tikzpicture}   
    \caption{Network architecture, showing encoding branch, parallel rotation branch (left), RNN bottleneck and decoder branch (right) with skip connections.}
    \label{fig:architecture}
\end{figure}

A convolutional recurrent neural network (CRNN) with a U-Net-style decoder maps the binaural and rotation features to the DirAC parameter spectrograms (Fig.~\ref{fig:architecture}). The spectral features are processed by a three-layer convolutional encoder with 128 channels, where strided convolutions progressively reduce the frequency resolution from 64 to 16 mel-bands while preserving the temporal resolution. Rotation features are processed on a parallel convolutional branch and concatenated with the spectral representation following \cite{garcia2022binaural}. The concatenated features are passed through a single-layer, unidirectional gated recurrent unit (GRU) with 256 hidden units to model temporal dependencies. The output is linearly projected and reshaped to a $128\times16$ representation, to match the encoder bottleneck. Unlike in \cite{garcia2022binaural}, this work requires spectrogram regression; to this end, a decoder branch progressively restores the frequency resolution to 64 mel-bands, with skip connections transferring fine-spectral information from corresponding encoder stages.
A final $1\times1$ convolution maps the decoder representation to the DirAC direction, represented as the Cartesian vector $\mathbf{d}(f,t)=[d_x,d_y,d_z]^T$, and diffuseness $\psi(f,t)$.

\subsubsection{Training}\label{sec:training}

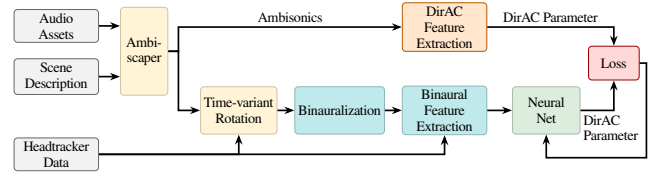
\begin{figure}[htbp]
    \centering
    \resizebox{1.01\columnwidth}{!}{\definecolor{blueColour}{HTML}{B0E3E6}
\definecolor{purpleColour}{HTML}{E1D5E7}
\definecolor{yellowColour}{HTML}{FFF2CC}
\definecolor{concatColor}{HTML}{999999}
\definecolor{greenColour}{HTML}{D5E8D4}

\begin{tikzpicture}[
    >={Stealth[scale=0.9]},
    thick,
    font=\large,
    node distance=5mm and 5mm,
    block/.style={draw, rounded corners=3pt, align=center, minimum width=5mm, minimum height=5mm, inner sep=4pt},
    input/.style={rounded corners=3pt, draw=black!60, fill=gray!10, minimum width=2.5cm, minimum height=0.8cm, align=center},
    yellowbox/.style={rounded corners=3pt,draw=yellowColour!80!black, fill=yellowColour!80, minimum width=1.5cm, minimum height=1.4cm, align=center},
    ambi/.style={rounded corners=3pt, draw=yellowColour!80!black, fill=yellowColour!80, minimum width=1.4cm, minimum height=2.6cm, align=center},
    orangebox/.style={rounded corners=3pt,draw=orange!70!black, fill=orange!20, minimum width=2.5cm, minimum height=1.4cm, align=center},
    bluebox/.style={rounded corners=3pt,draw=blueColour!80!black, fill=blueColour!80, minimum width=2.5cm, minimum height=1.4cm, align=center},
    dnnbox/.style={rounded corners=3pt,draw=greenColour!80!black, fill=greenColour!80, minimum width=2.0cm, minimum height=1.4cm, align=center},
    pinkbox/.style={rounded corners=3pt,draw=red!60!black, fill=red!15, minimum width=1.5cm, minimum height=1cm, align=center}
]

\node[input] (audio) at (0, 5) {Audio\\Assets};
\node[input] (scene) at (0, 3.5) {Scene\\Description};
\node[input] (head)  at (0, 1.2) {Headtracker\\Data};

\node[ambi] (ambi) at (2.6, 4.25) {Ambi-\\scaper};

\node[yellowbox] (rot)     at (5.4, 2.5) {Time-variant\\Rotation};
\node[bluebox, right=of rot]   (bin) {Binauralization};
\node[bluebox, right=of bin]   (binfeat) {Binaural\\Feature\\Extraction};

\node[orangebox, above=1cm of binfeat] (dirac) {DirAC\\Feature\\ Extraction};

\node[dnnbox] (dnn)  at (14.5, 2.5) {Neural\\Net};
\node[pinkbox]  (loss) at (16.5, 3.9) {Loss};


\draw[->, line width=0.5mm] (audio.east) -- (audio.east -| ambi.west);
\draw[->, line width=0.5mm] (scene.east) -- (scene.east -| ambi.west);

\draw[->, line width=0.5mm] (ambi.east) -- ++(0.3,0) |- (dirac.west) node[pos=0.75, above] {Ambisonics}; 
\draw[->, line width=0.5mm] (ambi.east) -- ++(0.3,0) |- (rot.west);

\draw[->, line width=0.5mm] (head.east) -- (rot.south |- head.east) -- (rot.south);
\draw[->, line width=0.5mm] (head.east) -| (binfeat.south);

\draw[->, line width=0.5mm] (dirac.east) -- (loss.north |- dirac.east) node[pos=0.5, above] {DirAC Parameter} -- (loss.north);

\draw[->, line width=0.5mm] (rot.east) -- (bin.west);
\draw[->, line width=0.5mm] (bin.east) -- (binfeat.west);
\draw[->, line width=0.5mm] (binfeat.east) -- (dnn.west);

\draw[->, line width=0.5mm] (dnn.east) -- node[pos=1.1, text width=2.0cm, below] {DirAC\\ Parameter} (loss.south |- dnn.east) -- (loss.south);

\draw[->, line width=0.5mm] (loss.east) -- ++(0.2,0) |- (dnn.south |- 0, 1.1) -- (dnn.south);

\end{tikzpicture}}
    \caption{Network training pipeline including dataset generation.}
    \label{fig:system}
\end{figure}

To train the neural network, ground-truth DirAC parameters are extracted 
as in \cite{pulkki2006directional,pulkkiFirstOrderDirAC2017}
from the ground-truth Ambisonics representation, see Fig.~\ref{fig:system}:
After STFT, for each TF bin $(f,t)$, the sound direction vector $\mathbf{d}(f,t)$ is computed from the active intensity and points towards the source:
\begin{equation}                      
\mathbf{d}(f,t) = \operatorname{Re}\!\left\{ W^*(f,t)[X(f,t),Y(f,t),Z(f,t)]^T \right\}, 
\label{eq:dirac_intensity}
\end{equation}
assuming SN3D normalization of the ambiX format.
The corresponding total acoustic energy density $E(f,t)$ is given by:
    \begin{equation}            
    E(f,t) = 0.5\left(|W(f,t)|^2 + |X(f,t)|^2 + |Y(f,t)|^2 + |Z(f,t)|^2\right)  
    \label{eq:dirac_energy}       
    \end{equation} 
The ground-truth diffuseness $\psi(f,t) \in [0, 1]$, describing the proportion of diffuse to total sound energy, is then calculated as:
\begin{equation}
    \psi(f,t) = 1 - \frac{\|\mathbf{d}(f,t)\|}{E(f,t) + \epsilon}.
    \label{eq:dirac_diffuseness}
\end{equation}

For an ideal single plane wave, $\|\mathbf{d}\| = E$, yielding $\psi = 0$, whereas for an isotropic diffuse field, $\mathbf{d} = \mathbf{0}$, yielding $\psi = 1$. 
In alignment with our proposed network architecture in Sec.~\ref{sec:model}, the DirAC parameters are processed using a 64-band Mel-scale.

To gain further insights into the model behavior, the network is trained with the dataset defined in Section~\ref{sec:evaluation} using either audio scenes without background noise (Dataset~A) or with varying levels of background noise (Dataset~B). Models were trained for 300 epochs using the AdamW optimizer with an initial learning rate of \(3\cdot 10^{-4}\) and weight decay of \(10^{-4}\). A cosine-annealing learning-rate schedule with warm restarts every 50 epochs and a minimum learning rate of \(10^{-6}\) was used.

\subsubsection{Loss Function}\label{sec:loss}
Inspired by the ACCDOA loss \cite{shimada2021accdoa}, we designed a loss function to jointly estimate source direction and diffuseness derived from the continuous DirAC parameters. 
To reduce contributions from overly diffuse time-frequency bins, where directional information is less perceptually relevant, we compute an activity-weighted target directional vector $\mathbf{d}_{\mathrm{acc}}$ as defined in Eq.~\ref{eq:loss_function}.
Here, the ground-truth Cartesian unit direction vector $\mathbf{u}$ is computed from the ground-truth intensity vector $\mathbf{d}$ and is weighted by the intensity magnitude $\|\mathbf{d}\|$ and its directionality estimated via (1-$\psi$), where $\psi$ is the ground-truth diffuseness.
\begin{equation}
a=\|\mathbf{d}\|(1-\psi),
\qquad
\mathbf{d}_{\mathrm{acc}}
=
a\mathbf{u}
=
\mathbf{d}(1-\psi).
\qquad
\label{eq:loss_function}
\end{equation}

In the resulting loss function, given in Eq.~\ref{eq:total_mae_loss}, the predicted Cartesian direction vector $\hat{\mathbf{d}}$ is compared with this activity-weighted target. Mean Absolute Error (MAE) is computed element-wise over the Cartesian components and TF bins. The parameter $\beta$ controls the contribution of the diffuseness term and was set to $0.2$. This formulation encourages the network to predict
stronger directional vectors for high-energy, directional TF bins while
reducing the contribution of diffuse regions. MSE was initially used for
both terms, but empirical evaluation showed that MAE provided improved resynthesis performance, hereafter denoted AW-MAE (MAE with activity-weighted directional target). 
\begin{equation}
    \mathcal{L}_{\text{AW-MAE}}
    =
    \operatorname{MAE}\;(\hat{\mathbf{d}},\;\mathbf{d}_{\mathrm{acc}})
    +
    \beta\;\;\operatorname{MAE}\;(\hat{\psi},\;\psi).
    \label{eq:total_mae_loss}
\end{equation}

\subsection{Ambisonics Synthesis}
The FOA signal $\mathbf{B}(f,t) = [W, Y, Z, X]^T$ is resynthesized in the STFT domain following DirAC principles \cite{pulkkiFirstOrderDirAC2017} 
from the estimated DirAC parameters. Because these parameters are estimated at mel-band resolution, they are mapped to the STFT frequency bins through nearest-neighbor interpolation.
As illustrated in Fig.~\ref{fig:inference}, the synthesis stage requires an omnidirectional sound pressure estimate $P(f,t)$, which must be approximated from the binaural signals, e.g., via complex averaging of $L(f,t)$ and $R(f,t)$:
\begin{equation}
    P(f,t) = 0.5\left(L(f,t) + R(f,t)\right).
    \label{eq:pressure_est}
\end{equation}
The sound pressure is subsequently decomposed into directional (direct) and diffuse components, $P_{\text{dir}}(f,t)$ and $P_{\text{diff}}(f,t)$, according to the predicted diffuseness $\hat{\psi}(f,t) \in [0, 1]$:
\begin{equation}
    P_{\text{dir}}(f,t) = \sqrt{1 - \hat{\psi}(f,t)}\,P(f,t),
    \label{eq:p_dir}
\end{equation}
\begin{equation}
    P_{\text{diff}}(f,t) = \sqrt{\hat{\psi}(f,t)}\,P(f,t).
    \label{eq:p_diff}
\end{equation}

The direct soundfield component is synthesized as a plane wave arriving from the predicted Cartesian unit direction $\hat{\mathbf{u}}$:
\begin{equation}
    \mathbf{B}_{\mathrm{dir}}(f,t)
    = P_{\mathrm{dir}}(f,t)
    [1,\hat{u}_y,\hat{u}_z,\hat{u}_x]^T,
    \label{eq:bdir}
\end{equation}
with $\hat{\mathbf{u}}(f,t)=\hat{\mathbf{d}}(f,t)/
(\|\hat{\mathbf{d}}(f,t)\|+\epsilon)$.
For the diffuse soundfield, $P_{\text{diff}}(f,t)$ is processed with mutually orthogonal decorrelators $H_m(f)$ to generate four uncorrelated B-format components $\mathbf{B}_{\text{diff}}(f,t)$, preserving energy and reproducing an isotropic soundfield \cite{pulkkiFirstOrderDirAC2017}. The complete FOA representation is obtained by combining the direct and diffuse streams:
\begin{equation}
    \mathbf{B}(f,t) = \mathbf{B}_{\text{dir}}(f,t) + \mathbf{B}_{\text{diff}}(f,t).
    \label{eq:b_total}
\end{equation}
Finally, the time-domain FOA signal is reconstructed by applying the inverse STFT with synthesis windowing and overlap-add.

\section{Evaluation}
\label{sec:evaluation}

\subsection{Dataset Generation} \label{sec:dataset_generation}
To train the Bin2Ambi architecture, a large-scale dataset that consists of time-aligned Ambisonics and head-tracked binaural audio is required. Training data was generated using the pipeline shown in Fig.~\ref{fig:system}. 
First, static 5th-Order Ambisonics sound scenes were generated at \SI{48}{kHz} using Ambiscaper \cite{perez2018ambiscaper}.
Speech, music, and sound events from the datasets NIGENS \cite{trowitzsch2019nigens}, DESRA \cite{gygi2010development}, and DataSEC \cite{fredianelli2025environmental} were spatialized as foreground sounds at random directions and gain levels. Short sound events were stitched with recordings from the same sound class to reduce scene sparsity, following \cite{garcia2022binaural}. Source recordings were partitioned before scene generation to ensure disjoint training, validation, and test material. Optionally, environmental background noise in Higher-Order Ambisonics was added from ARTE \cite{buchholz_2019_3386569} and the 3rd Clarity Challenge \cite{11461465,clarity_task3_data} at SNRs between -5 and \SI{30}{dB}.
Each scene has a duration of \SI{8}{\second} and consists of up to five foreground sounds. In total, we generated 12,000 scenes (i.e. more than \SI{26.5}{\hour} total duration), 50\% without background noise (defined as Dataset A) and 50\% with background noise (Dataset B). 

The scenes were intentionally generated in 5th-Order Ambisonics to derive binaural audio samples of high spatial accuracy via convolution with \mbox{KU-100} HRTFs in the SH domain. To simulate time-varying head movements, soundfields were rotated prior binauralization using 3 DoF sensor data from a Supperware Head~Tracker~1 at \SI{100}{Hz} update rate.

\subsection{Tested Conditions}
Since we introduce the Bin2Ambi task in this contribution, no directly comparable baseline system is yet available. Thus, we assess four differently trained models against the FOA ground-truth.
To account for the inherent limitations of DirAC \cite{pulkki2006directional}, we also include
DirAC oracle parameters that were directly derived from the ground-truth FOA.

All conditions are summarized in Tab.\ref{tab:conditions}.
Binaural input features and DirAC parameters were computed every \SI{10.6}{ms}. Objective metrics are computed from both the DirAC parameters and from the resynthesized FOA signals.
For a subjective listening test, the FOA signals are rendered to a 9.1.4 loudspeaker setup using the AllRAD method \cite{zotter2012all}.

\begin{table}[htbp]    
    \footnotesize
    \centering
    \caption{Evaluated Conditions}
    \label{tab:conditions}    
    \begin{tabular}{l l}    
        \toprule
        Condition & Description\\
        \midrule
        REF & FOA Reference\\
        DIRAC GT  & The DirAC oracle parameter. The upper bound.\\
        AW-MAE & Model trained with AW-MAE Loss (Eq. \ref{eq:total_mae_loss}) on Dataset A\\
        AW-MSE & Model trained with AW-MSE Loss on Dataset A\\ 
        AW-MSE BG & Model trained with AW-MSE Loss on Dataset B\\
        Untrained & System with randomly initialised model weights\\
        \bottomrule
    \end{tabular}    
\end{table}

\subsection{Results}
\subsubsection{Objective Metrics}
The Mean-Squared Error (MSE) between the estimated and ground-truth DirAC parameters is used to measure regression accuracy. To assess directional accuracy, we defined the weighted mean angular error (WMAE) as:
\begin{equation}
    \mathrm{WMAE} =
    \frac{\sum_{f,t}\theta(f,t)\|\mathbf{v}(f,t)\|}
         {\sum_{f,t}\|\mathbf{v}(f,t)\|+\epsilon},
    \label{eq:wmae}
\end{equation}
where $\theta(f,t)$ is the angular error between the predicted and
ground-truth intensity vectors, calculated using cosine similarity,
and $\mathbf{v}(f,t)$ is the ground-truth intensity vector. Weighting by $|\mathbf{v}(f,t)|$ emphasizes TF bins containing greater directional energy.

\begin{figure}[!t]
    \centering
    \hspace*{-6mm}
    \footnotesize
    \includegraphics[width=0.99\columnwidth]{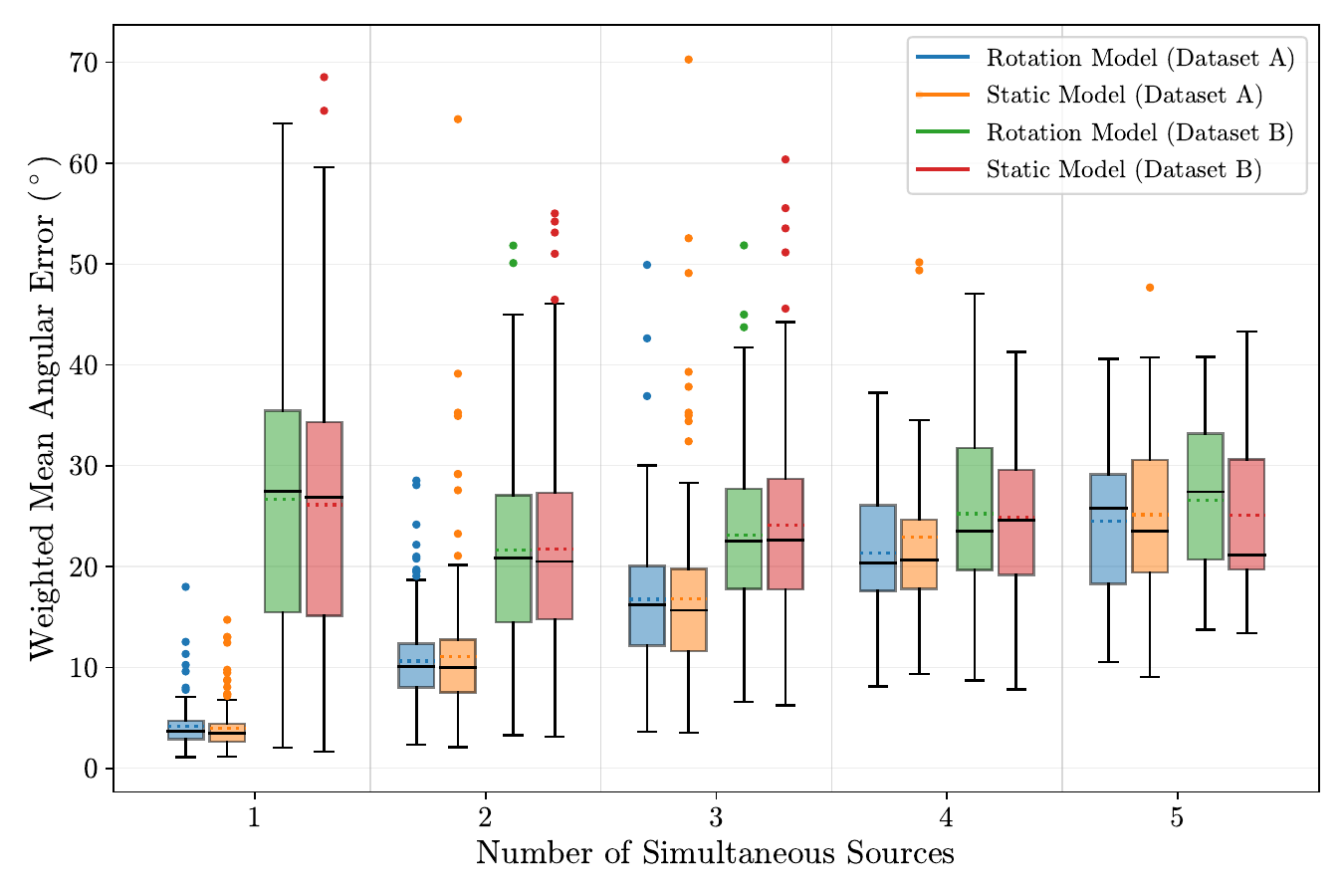}
    \caption{WMAE distributions for different numbers of simultaneous sources in a scene for models trained with the AW-MAE loss.}
    \label{fig:wmae_boxplot}
\end{figure}

The WMAE for scenes with varying numbers of sources is shown in Fig.~\ref{fig:wmae_boxplot}. Directional accuracy decreased as the number of simultaneous sources increased. The WMAE was noticeably higher in the presence of background noise (Dataset B). We did not find substantial improvement in mean WMAE due to rotation features; however, as visible in Fig.\ref{fig:wmae_boxplot}, extreme directional errors are prevented for the rotation model compared with the static model. Consistent with this observation, the 99th percentile WMAE decreased from $49.1^\circ$ to $35.4^\circ$
in Dataset A (without background noise). This suggests that the main benefit of head rotation is in reducing large localization errors (i.e., front-back confusion) rather than uniformly improving directional accuracy. This extends the findings of \cite{garcia2022binaural}, where head rotation was shown to improve binaural DOA estimation.

\begin{figure}[h]
    \centering
    \hspace*{-6mm}
    \includegraphics[width=0.99\columnwidth]{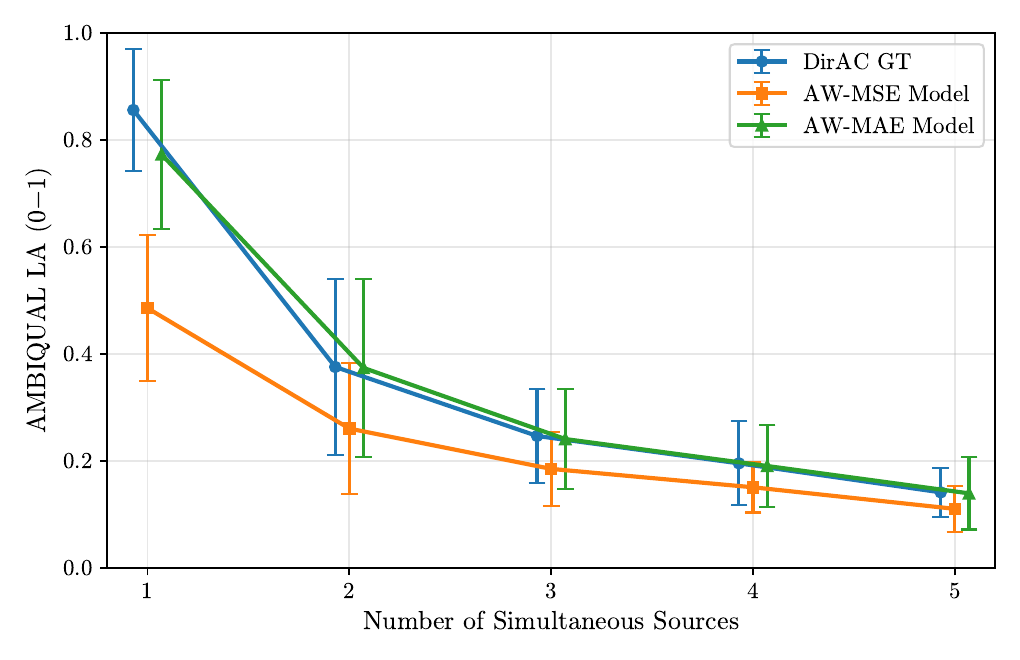}
    \caption{AMBIQUAL Localization Accuracy as a function of scene complexity in Dataset A (Mean $\pm$ Std.).}
    \label{fig:listening_accuracy}
    \vspace{-2mm}
\end{figure}

The resynthesized FOA signals were also evaluated using AMBIQUAL \cite{narbutt2018ambiqual}. In particular, the AMBIQUAL Localization Accuracy (LA) is used to objectively assess whether improvements in the DirAC parameters translated to improved spatial localization after resynthesis. Fig.~\ref{fig:listening_accuracy} shows the AMBIQUAL LA as a function of the number of events in the scene for the no-background condition and for the AW-MSE and AW-MAE loss functions. The AW-MAE variant of the model followed the resynthesized DirAC ground-truth more closely than the AW-MSE variant, particularly for sparse scenes with few independent sources. This indicates that similar parameter-domain errors do not necessarily translate to equivalent localization accuracy after resynthesis. 
With increasing scene complexity, overlapping sound events cause the diffusion parameter $\psi$ to trend higher, resulting in more decorrelated sound energy in the FOA output. This leads to a sharp decrease in AMBIQUAL's LA measure, which is very sensitive to phase differences.

\subsubsection{Ablation Study}\label{sec:ablation}

An ablation study was performed using Dataset A's test set of 600 items. A lightweight CNN, consisting of the encoder and rotation branch without the recurrent or decoder components, is used as a naive baseline. Introducing the recurrent component substantially reduced WMAE from $40.25^\circ$ to $13.87^\circ$, providing the largest architectural improvement. Also clearly visible is the low model performance without head tracking. 
Subsequent architectural and loss-function changes provided smaller gains, with the lowest WMAE of $11.82^\circ$ obtained using the AW-MAE loss (see Tab.~\ref{tab:ablation}).
\vspace{-0.15cm}
\begin{table}[htbp]
    \centering
    \footnotesize
    \caption{Ablation study.}
    \label{tab:ablation}
    \setlength{\tabcolsep}{3.2pt}
    \renewcommand{\arraystretch}{1.05}

    \resizebox{\columnwidth}{!}{%
    \begin{tabular}{lccccc}
        \toprule
        \textbf{Variant} &
        \textbf{Params} &
        \textbf{MSE} $\downarrow$ &
        \textbf{WMAE ($^\circ$)} $\downarrow$ &
        \textbf{P99 ($^\circ$)} $\downarrow$ &
        \textbf{Max ($^\circ$)} $\downarrow$ \\
        \midrule

        Simple CNN
        & 319 K & 1.26e-01 & 40.25 & 68.89 & 83.74 \\

        No Decoder
        & 2.25 M & 3.63e-02 & 13.87 & 40.17 & 59.74 \\

        No Skip Conn.
        & 2.88 M & 3.29e-02 & 12.16 & 36.85 & 52.01 \\

        Standard MSE Loss
        & 2.98 M & \textbf{3.05e-02} & 12.23 & \textbf{34.20} & 54.07 \\

        No Head Tracking (Static) 
        & 2.86 M & 3.46e-02 & 12.35 & 49.10 & 70.28 \\

        AW-MSE Loss
        & 2.98 M & 3.18e-02 & 12.05 & 35.35 & 49.92 \\

        AW-MAE Loss
        & 2.98 M & 3.33e-02 & \textbf{11.82} & 37.11 & \textbf{48.06} \\

        \bottomrule
    \end{tabular}%
    }
\end{table}
\vspace{-0.5cm}
\subsubsection{Subjective Test}
\begin{figure}[htbp]
    \centering
    \vspace{-0.4cm}
    \resizebox{0.7\columnwidth}{!}{\input{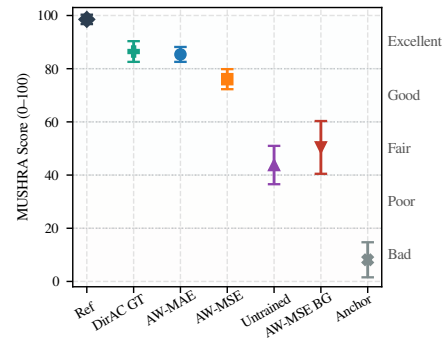}}
    \vspace{-0.5cm}
    \caption{MUSHRA scores (Mean $\pm$ 95\% CI) pooled across test items.}
    \label{fig:mushra}
    \vspace{-4mm}
\end{figure}

Nine subjects participated in a MUSHRA listening test \cite{bs20141534}, with the hidden reference and a \SI{3.5}{kHz} low-pass anchor present in every trial.
Participants were asked to ``Rate the spatial quality of the system under test compared to the reference'' for ten different test items. These items contain up to four spatially distributed musical stems from \mbox{MUSDB18-HQ} \cite{musdb18-hq} and different background noise at varying SNRs.
Figure~\ref{fig:mushra} shows the pooled results after post-screening with seven valid listeners. 
Among the evaluated conditions, DirAC GT was rated highest with 86.5 points, closely followed by the proposed AW-MAE model with 85.4 points. The Untrained condition received the lowest score.
Paired comparison tests with Holm-Bonferroni correction suggest no statistically significant differences ($\rho=.05$) between DirAC GT and AW-MAE, indicating that our proposed model can reliably estimate the DirAC parameters from binaural audio.

\section{Conclusion}
\label{sec:summary}
We introduced the novel task of Binaural to Ambisonics conversion (Bin2Ambi) and proposed a CRNN with a U-Net-style decoder to predict DirAC parameters from binaural audio features. The results demonstrate the feasibility of using predicted DirAC parameters for Bin2Ambi, with directional accuracy dependent on scene complexity and the ambient background noise level. Head rotation was shown to reduce extreme directional errors and a perceptual test found that the proposed system achieved spatial quality comparable to the DirAC oracle. Future work will focus on better estimating the omnidirectional component from the binaural audio input and optimizing the system for more varied acoustic conditions.

\balance
\bibliographystyle{IEEEbib}
\bibliography{bibliography}

@book{blauert1996spatial,
  title={Spatial hearing: the psychophysics of human sound localization},
  author={Blauert, Jens},
  year={1996},
  publisher={The MIT press}
}

@incollection{pulkkiFirstOrderDirAC2017,
  author    = {Pulkki, Ville and Politis, Archontis and Laitinen, Mikko-Ville and Vilkamo, Juha and Ahonen, Jukka},
  title     = {First-Order Directional Audio Coding ({DirAC})},
  booktitle = {Parametric Time-Frequency Domain Spatial Audio},
  publisher = {Wiley},
  year      = {2017},
  pages     = {89--140},
  doi       = {10.1002/9781119252634.ch5},
}

@article{olivieri2019scene,
  title={Scene-based audio and higher order ambisonics: A technology overview and application to next-generation audio, {VR} and 360 video},
  author={Olivieri, Ferdinando and Peters, Nils and Sen, Deep},
  journal={EBU Tech},
  year={2019}
}

@inproceedings{garcia2022binaural,
  title={Binaural source localization using deep learning and head rotation information},
  author={Garc{\'\i}a-Barrios, Guillermo and Krause, Daniel Aleksander and Politis, Archontis and Mesaros, Annamaria and Guti{\'e}rrez-Arriola, Juana M and Fraile, Rub{\'e}n},
  booktitle={30th European Signal Processing Conference (EUSIPCO)},
  pages={36--40},
  year={2022},
  organization={IEEE}
}

@INPROCEEDINGS{narbutt2018ambiqual,
  author={Narbutt, Miroslaw and Allen, Andrew and Skoglund, Jan and Chinen, Michael and Hines, Andrew},
  booktitle={10th International Conference on Quality of Multimedia Experience (QoMEX)}, 
  title={AMBIQUAL - a full reference objective quality metric for ambisonic spatial audio}, 
  year={2018},  
  doi={10.1109/QoMEX.2018.8463408}}

@inproceedings{perez2018ambiscaper,
  title={Ambiscaper: A tool for automatic generation and annotation of reverberant ambisonics sound scenes},
  author={Perez-Lopez, Andres},
  booktitle={2018 16th International Workshop on Acoustic Signal Enhancement (IWAENC)},
  pages={1--9},
  year={2018},
  organization={IEEE}
}

@inproceedings{pulkki2006directional,
  title={Directional audio coding in spatial sound reproduction and stereo upmixing},
  author={Pulkki, Ville},
  booktitle={Proc. of the AES 28th Int. Conf, Pitea, Sweden},
  year={2006}
}

@inproceedings{shimada2021accdoa,
  title={ACCDOA: Activity-coupled cartesian direction of arrival representation for sound event localization and detection},
  author={Shimada, Kazuki and Koyama, Yuichiro and Takahashi, Naoya and Takahashi, Shusuke and Mitsufuji, Yuki},
  booktitle={IEEE international conference on acoustics, speech and signal processing (ICASSP)},
  pages={915--919},
  year={2021},
  organization={IEEE}
}

@article{fredianelli2025environmental,
  title={Environmental noise dataset for sound event classification and detection},
  author={Fredianelli, Luca and Artuso, Francesco and Pompei, Geremia and Licitra, Gaetano and Iannace, Gino and Akbaba, Andac},
  journal={Scientific Data},
  volume={12},
  number={1},  
  year={2025}  
}

@article{gygi2010development,
  title={Development of the database for environmental sound research and application ({DESRA}): Design, functionality, and retrieval considerations},
  author={Gygi, Brian and Shafiro, Valeriy},
  journal={EURASIP Journal on Audio, Speech, and Music Processing},
  volume={2010},
  number={1},
  pages={654914},
  year={2010}  
}

@article{trowitzsch2019nigens,
  title={The {NIGENS} general sound events database},
  author={Trowitzsch, Ivo and Taghia, Jalil and Kashef, Youssef and Obermayer, Klaus},
  journal={arXiv preprint arXiv:1902.08314},
  year={2019}
}

@INPROCEEDINGS{11461465,
  author={Barker, Jon and Akeroyd, Michael A. and Cox, Trevor J. and Culling, John F. and Firth, Jennifer and Graetzer, Simone and Naylor, Graham},
  booktitle={IEEE International Conference on Acoustics, Speech and Signal Processing (ICASSP)}, 
  title={The 3rd Clarity Prediction Challenge: A Machine Learning Challenge for Hearing aid Speech Intelligibility Prediction}, 
  year={2026},
  volume={},
  number={},    
  doi={10.1109/ICASSP55912.2026.11461465}}

@dataset{clarity_task3_data,
  author={Barker, Jon and Akeroyd, Michael A. and Cox, Trevor J. and Culling, John F. and Firth, Jennifer and Graetzer, Simone and Naylor, Graham},
  title        = {{3rd Clarity Enhancement Challenge - Task 3 Dataset}},  
  year         = 2025,    
  note          = {\url{https://claritychallenge.org/docs/cec3/task_3/cec3_task3_data}},
}

@article{buchholz_2019_3386569,
  title={The ambisonic recordings of typical environments ({ARTE}) database},
  author={Weisser, Adam and Buchholz, J{\"o}rg M and Oreinos, Christos and Badajoz-Davila, Javier and Galloway, James and Beechey, Timothy and Keidser, Gitte},
  journal={Acta Acustica United With Acustica},
  volume={105},
  number={4},
  pages={695--713},
  year={2019},
  publisher={S. Hirzel Verlag GmbH}
}

@inproceedings{gebruA2BNeuralRendering2025,
  title = {{{A2B}}: {{Neural Rendering}} of {{Ambisonic Recordings}} to {{Binaural}}},
  shorttitle = {{{A2B}}},
  booktitle = {{{IEEE International Conference}} on {{Acoustics}}, {{Speech}} and {{Signal Processing}} ({{ICASSP}})},
  author = {Gebru, Israel D. and Keebler, Todd and Sandakly, Jake and Krenn, Steven and Markovi{\'c}, Dejan and Buffalini, Julia and Hassel, Samuel and Richard, Alexander},
  year = 2025,
  address = {Hyderabad, India},  
  pages = {1--5},
  issn = {2379-190X},
  doi = {10.1109/ICASSP49660.2025.10890507},
  urldate = {2025-10-07}
}

@article{yinEndtoEndPairedAmbisonicBinaural2024,
  title = {End-to-{{End Paired Ambisonic-Binaural Audio Rendering}}},
  author = {Yin, Zhu and Kong, Qiuqiang and Shi, Junjie and Liu, Shilei and Ye, Xuzhou and Wang, Ju-Chiang and Shan, Hongming and Zhang, Junping},
  year = 2024,  
  journal = {IEEE/CAA Journal of Automatica Sinica},
  volume = {11},
  number = {2},
  pages = {502--513},
  issn = {2329-9274},
  doi = {10.1109/JAS.2023.123969},
  urldate = {2025-10-07}
}

@inproceedings{Turner2024TFSpatialUpmix,
  author    = {Daniel Turner and Damian T. Murphy},
  title     = {A Deep Learning Approach to the Prediction of Time-Frequency Spatial Parameters for Use in Stereo Upmixing},
  booktitle = {Proceedings of the 27th International Conference on Digital Audio Effects (DAFx24)},
  year      = {2024},
  pages     = {428--435},
  address   = {Guildford, Surrey, UK}  
}

@article{adavanneSoundEventLocalization2019,
  title = {Sound {{Event Localization}} and {{Detection}} of {{Overlapping Sources Using Convolutional Recurrent Neural Networks}}},
  author = {Adavanne, Sharath and Politis, Archontis and Nikunen, Joonas and Virtanen, Tuomas},
  year = 2019,  
  journal = {IEEE Journal of Selected Topics in Signal Processing},
  volume = {13},
  number = {1},
  pages = {34--48},
  issn = {1941-0484},
  doi = {10.1109/JSTSP.2018.2885636},
  urldate = {2025-11-04}
}

@inproceedings{fleischhauerFullSphereBinauralDirectionofArrival2025,
  author = {Fleischhauer, Erik and Jax, Peter},
  title = {Full-sphere binaural direction-of-arrival estimation incorporating head rotation information},
  booktitle = {European Signal Processing Conference (EUSIPCO)},
  address = {Palermo, Italy},  
  year = {2025},
  pages = {231--235}
}

@misc{musdb18-hq,
  author       = {Rafii, Zafar and
                  Liutkus, Antoine and
                  Stöter, Fabian-Robert and
                  Mimilakis, Stylianos Ioannis and
                  Bittner, Rachel},
  title        = {{MUSDB18-HQ} - an uncompressed version of {MUSDB18}},  
  year         = 2019,
  doi          = {10.5281/zenodo.3338373},
  note          = {\url{https://doi.org/10.5281/zenodo.3338373}}
}

@article{zotter2012all,
  title={All-round ambisonic panning and decoding},
  author={Zotter, Franz and Frank, Matthias},
  journal={Journal of the Audio Engineering Society},
  volume={60},
  number={10},
  pages={807--820},
  year={2012}
}

@techreport{bs20141534,
type = {Standard},
author       = {{ITU-R BS.1534-2}},
key = {ITU-R BS.1534-2},
year = {2014},
title = {Method for the subjective assessment of intermediate quality level of audio systems},
volume = {2014},
address = {Geneva, CH},
institution = {International Telecommunication Union}
}

@inproceedings{liang2026immersiveflow,
  title={ImmersiveFlow: Stereo-to-7.1.4 spatial audio generation with flow matching},
  author={Liang, Zining and Wang, Runbang and Ye, Xuzhou and Kong, Qiuqiang},
  booktitle = {9th International Workshop on Acoustic Signal Enhancement (IWAENC)},
  address = {Cremona, Italy},    
  year={2026}
}

@ARTICLE{MPEG-H_2017,
  author={Bleidt, Robert L. and others},
  journal={IEEE Transactions on Broadcasting},
  title={Development of the {MPEG-H TV} Audio System for {ATSC} 3.0},
  year={2017},
  volume={63},
  number={1},
  pages={202-236},
  doi={10.1109/TBC.2017.2661258}}

@inproceedings{multrus2024immersive,
  title={Immersive Voice and Audio Services ({IVAS}) codec - The new {3GPP} standard for immersive communication},
  author={Multrus, Markus and Bruhn, Stefan and others},
  booktitle={157th AES Convention},
  address = {Long Beach, CA, US},
  year={2024}
}

\end{document}